\documentclass[final,authoryear,5p,times,twocolumn]{elsarticle}

\usepackage{amssymb}

\biboptions{square,comma,numbers,sort&compress}

\journal{Physica Scripta}

\usepackage{epsfig}
\usepackage{psfrag}
\newcommand{\mypsfrag}[2]{\psfrag{#1}{\footnotesize{#2}}}
\usepackage{amsfonts}          
\usepackage{amsmath}           
\usepackage[a4paper,ps2pdf,colorlinks=true,breaklinks=true]{hyperref}
\usepackage{ulem}
\usepackage{xspace}

\renewcommand{\vec}[1]{\mbox{\boldmath $ #1$}}

\newcommand{\x}{\cdot}

\newcommand{\dd}{\partial}
\newcommand{\de}{{\rm \, d}}

\providecommand{\R}{\ensuremath{R}\xspace}
\renewcommand{\P} {\ensuremath{P}\xspace}
\providecommand{\Pm}{\ensuremath{P_m}\xspace}
\providecommand{\ttau}{\ensuremath{\tau}\xspace}

\providecommand{\MD}{\textbf{MD}\xspace}
\providecommand{\FD}{\textbf{FD}\xspace}
\newcommand{\tA}{$\tau_\mathrm{MD}$}
\newcommand{\tB}{$\tau_\mathrm{FD}$}

\providecommand{\Mdip}{\ensuremath{{M}_\text{dip}}\xspace}

\providecommand{\Efp}{\ensuremath{\widetilde{E}_p}\xspace}
\providecommand{\Emt}{\ensuremath{\overline{E}_t}\xspace}
\providecommand{\Eft}{\ensuremath{\widetilde{E}_t}\xspace}
\providecommand{\Ha}{\ensuremath{H_{\vec{_a}}}\xspace}
\providecommand{\Hu}{\ensuremath{H_{\vec{_u}}}\xspace}
\providecommand{\HB}{\ensuremath{H_{\vec{_B}}}\xspace}
\providecommand{\Hcross}{\ensuremath{H_\times}\xspace}
\providecommand{\MfpToMmp}{\ensuremath{\widetilde{M}_p/\overline{M}_p}\xspace}

\usepackage{color}

\newcommand{\revb}[1]{}

\newcommand{\hide}[1]{}

\allowdisplaybreaks

\newlength{\shortwidth}
\newlength{\textwidtha}
\begin{document}

\begin{frontmatter}



\title{Bistable attractors in a model of \\ 
       convection-driven spherical dynamos}


\author{\textbf{Radostin D~Simitev}$^{1,3}$ \textbf{and Friedrich H Busse}$^{2,3}$}
\address{$^1$ School of Mathematics and Statistics, University of
  Glasgow -- Glasgow G12 8QW, UK, EU\\
$^2$ Institute of Physics,  University of Bayreuth -- Bayreuth
  D-95440, Germany, EU\\
$^3$ NORDITA, AlbaNova University Center -- Stockholm SE-10691,
  Sweden, EU\\[2mm]
E-mail: \href{mailto:Radostin.Simitev@glasgow.ac.uk}{Radostin.Simitev@glasgow.ac.uk}
}

\begin{abstract}
The range of existence and the properties of two essentially different chaotic
attractors found in a model of nonlinear convection-driven dynamos
in rotating spherical shells are investigated. 
A hysteretic transition between these attractors is established as a
function of the rotation parameter \ttau.
The width of the basins of attraction is also estimated. 
\end{abstract}


\end{frontmatter}

\section{Introduction}

Chaotic systems by virtue of their apparently ``random'' fluctuations,
are expected to frequent, with little bias, all points in their phase 
space over wide parameter ranges. This view implies that there
should not be any abrupt transitions between distinguishable
chaotic states. Indeed, non-linear transitions between attractors
are rarely found in turbulent fluids. Contrary to this expectation, a
number of examples of  discontinuous transitions have been observed
recently in chaotic and turbulent fluid systems
\cite{Ravelet2004,Mujica2006,Berhanu2009,Weiss2010}. 
Bifurcations between distinguishable chaotic states appear to be a little less
unusual in magnetohydrodynamic flows because of the additional degrees
of freedom offered by the magnetic  field. For instance, various types
of dipolar, quadrupolar, hemispherical dynamos, and bifurcations
between them are routinely reported in numerical simulations,
e.g. \cite{Simitev2005Prandtlnumber,Busse2006Parameter}. 
In turn, dipolar dynamos are typically found to belong to two distinct
regimes -- a regime with strong dipolar field, and another regime with
weaker dipolar component and significant multipole contributions
e.g. \cite{Busse2006Parameter,Christensen2006Scaling}.  

Far more remarkable is a recent finding that two essentially
different chaotic dipolar dynamo solutions may exist at identical
values of the basic parameters of a generic model of convection-driven
dynamos in rotating  spherical shells \cite{Simitev2009}.   
Such bistability offers the possibility of a hysteretic transition.
While hysteresis was established in \cite{Simitev2009} as a function
of three of the parameters in the problem -- the Rayleigh, the
ordinary, and the magnetic Prandtl numbers, to be defined below, the
dependence on the last remaining basic parameter, the 
Coriolis number $\ttau$, was not studied there. This, however, leaves an
important gap since the variation in $\ttau$ in this minimal
self-consistent model of spherical convective dynamos describes the
different rotation rates characteristic for various cosmic objects. In
addition, given 
that current geodynamo simulations are unable to achieve geophysically
realistic values of $\ttau$, extrapolation of the $\ttau$
dependence is heavily used to compare models and observations.
As a partial remedy, bistability has been demonstrated for two specific
values of $\ttau$ in \cite{Busse2010}. For these reasons, here we wish
to investigate the full extent of the coexistence and hysteresis in
dependence of the rotation parameter $\ttau$. A number of other
important questions left open in \cite{Simitev2009,Busse2010},
including the width of the basins of attraction of the distinguishable
chaotic states and the possibility of spontaneous transition will also
be discussed in the present 
paper, along with results on essential properties of dynamo
action such as kinetic, magnetic and cross-helicity
generation. 



\section{Formulation and methods}

\subsection{Model}
We employ a minimal model of nonlinear convection-driven dynamo
process in rotating spherical shells with the aim to retain general
applicability of the results to a variety of cosmic objects and to
understand fundamental physical mechanisms. The model is identical to the one described in 
\cite{Simitev2009} but for completeness we provide a concise formulation
below. We consider a spherical fluid shell of thickness
$d$ rotating with a constant angular velocity $\Omega$. The existence of a static state is
assumed with {a} temperature distribution $T_S = T_0 - \beta d^2 r^2 /2$
and a gravity field in the form $\vec g = - d \gamma \vec r$, where
$rd$ is the length of the position vector with respect to the center
of the sphere. {This form of temperature profile alludes
to the possibility that at least a fraction of the energy available to
planetary dynamos is due to radiogenic heat release.}
In addition to  $d$, we use the time $d^2 / \nu$,  the temperature
$\nu^2 / \gamma \alpha d^4$ and 
the magnetic flux density $\nu ( \mu \varrho )^{1/2} /d$ as scales for
the dimensionless description of the problem  where $\nu$ denotes the
kinematic viscosity of the fluid, $\kappa$ its thermal diffusivity, 
$\varrho$ its density and $\mu$ its magnetic permeability.
In common with most other simulations of Earth and planetary
dynamos \cite{DormyNumerical,Kono2002Recent},
we assume the Boussinesq approximation implying a constant density
$\varrho$ except in the gravity term where its temperature
dependence is taken into account with $\alpha \equiv - (
\de\varrho/\de T)/\varrho =$ const. 
The equations of motion for the velocity vector $\vec u$, the heat
equation for the deviation  $\Theta$ from the static temperature 
distribution, and the equation of induction for the magnetic flux
density $\vec B$ are then given by  
\begin{subequations}
\begin{gather}
\label{1b}
\nabla \cdot \vec u = 0, \qquad \nabla \cdot \vec B = 0, \\
\label{1a} 
(\partial_t + \vec u \cdot \nabla )\vec u + \tau \vec k \times
\vec u = - \nabla \pi +\Theta \vec r + \nabla^2 \vec u + \vec B \cdot
\nabla \vec B, \\
\label{1c}
P(\partial_t \Theta + \vec u \cdot \nabla \Theta) = R \vec r \cdot \vec u + \nabla^2 \Theta, \\
\label{1d} 
\nabla^2 \vec B =  P_m(\partial_t \vec B + \vec u \cdot \nabla \vec B
-  \vec B \cdot \nabla \vec u),
\end{gather}
\end{subequations}
where all gradient terms in the equation of motion 
have been combined into $ \nabla \pi$. The dimensionless
parameters in our formulation are the Rayleigh number $R$, the
Coriolis number $\tau$, the Prandtl number $P$ and the magnetic
Prandtl number $P_m$,  
\begin{equation}
R = \frac{\alpha \gamma \beta d^6}{\nu \kappa} , 
\enspace \tau = \frac{2
\Omega d^2}{\nu} , \enspace P = \frac{\nu}{\kappa} , \enspace P_m = \frac{\nu}{\lambda},
\end{equation}
where $\lambda$ is the magnetic diffusivity.  Being solenoidal vector
fields,  $\vec u$ and $\vec B$ can be represented uniquely in terms
of poloidal and toroidal components,
\begin{subequations}
\begin{gather}
\vec u = \nabla \times ( \nabla v \times \vec r) + \nabla w \times 
\vec r \enspace , \\
\vec B = \nabla \times  ( \nabla h \times \vec r) + \nabla g \times 
\vec r \enspace .
\end{gather}
\end{subequations}
We assume fixed temperatures at $r=r_i \equiv 2/3$ and  $r=r_o \equiv
5/3$ and stress-free {rather than no-slip boundary conditions in order
to approach, at least to some extent, the extremely low values of viscosity
believed to be appropriate to planetary cores
\cite{Kuang1997Earthlike}},
\begin{equation}
\label{vbc}
\hspace*{-8mm}
v = \partial^2_{rr}v = \partial_r (w/r) = \Theta = 0.
\end{equation}
Two conditions on the poloidal scalar $v$ are required at each
boundary because the corresponding poloidal equation is obtained by
taking $\vec r\cdot\nabla\times\nabla\times$~ of (1b) and thus it is
of higher order as discussed below. 
For the magnetic field we
assume electrically insulating boundaries at $r=r_i$ and  $r=r_o$ such
that the poloidal function $h$ matches the function $h^{(e)}$ which
describes the potential fields  outside the fluid shell,  
\begin{equation}
\hspace*{-8mm}
\label{mbc}
g = h-h^{(e)} = \partial_r ( h-h^{(e)})=0\;\; \mbox{at}\;\; r=r_i, r_o.
\end{equation}
The radius ratio $r_i/r_o = 0.4$ is slightly larger than that
appropriate for the Earth's liquid core. 
This is a standard formulation of the spherical convection-driven
dynamo problem 
\cite{DormyNumerical,Busse2000Homogeneous,Kono2002Recent}
for which an extensive  collection of results already exists
\cite{Grote2000Regular,GroBu,Simitev2005Prandtlnumber,Busse2006Parameter}. 
The results reported below are not strongly model dependent. In
particular, dynamos with stress-free and with no-slip velocity boundary
conditions as well as with different modes of energy supply are known to
have comparable energy densities and symmetry properties (see fig.~15 of
\cite{Kono2002Recent}). 

\subsection{Methods of solution}
Equations of motion for the scalar fields $v$, $w$, are obtained by
taking $\vec r\cdot\nabla\times\nabla\times$~ and $\vec
r\cdot\nabla\times$~   of equation \eqref{1a} and
equations for $g$ and $h$ are obtained by taking $\vec
r\cdot\nabla\times$~ and $\vec r\;\cdot$~  of equation \eqref{1d}. 
These equations are solved numerically by a 
pseudo-spectral method as described in \cite{Tilgner1999Spectral} 
based on expansions of all dependent variables in
spherical harmonics for the angular dependences and in Chebychev
polynomials for the radial dependence. 
A minimum of 41
collocation 
points in the radial direction and spherical harmonics up to the
order 96 have been used in all cases reported here which provides
adequate resolution 
The dynamo solutions are characterized by their magnetic energy
densities, 
\begin{gather}
\overline{M}_p = \frac{1}{2} \langle \mid\nabla \times ( \nabla \overline{h}
\times \vec r )\mid^2 \rangle ,  \quad
 \overline{M}_t = \frac{1}{2} \langle \mid\nabla
\overline g \times \vec r \mid^2 \rangle, \nonumber\\
\widetilde{M}_p = \frac{1}{2} \langle \mid\nabla \times ( \nabla \widetilde h
\times \vec r )\mid^2 \rangle , \quad 
 \widetilde{M}_t = \frac{1}{2} \langle \mid\nabla
\widetilde g \times
\vec r \mid^2 \rangle, \nonumber
\end{gather}
where $\langle\cdot\rangle$ indicates the average over the fluid shell
and $\overline h$ refers to the axisymmetric component of $h$,
while $\widetilde h$ is defined by $\widetilde h = h - \overline h $. 
The corresponding kinetic energy densities $\overline{E}_p$,
$\overline{E}_t$, $\widetilde{E}_p$ and $\widetilde{E}_t$
are defined analogously with $v$ and $w$ replacing $h$ and
$g$. The total magnetic energy density is
$M=\overline{M}_p+\overline{M}_t+\widetilde{M}_p+\widetilde{M}_t$, and
similarly for the total kinetic energy density $E$.
In addition, the magnetic energy densities can be divided into those of
fields that are antisymmetric (axial dipole symmetry) and those that
are symmetric (axial quadrupole symmetry) with respect to the
equatorial plane. The former (latter) are described  by spherical
harmonics $Y_l^m$ with odd (even) $l+m$. 
Other quantities of interest are the helicity density of a vector field
$\vec{a}$, 
$$
\Ha = \vec{a} \cdot (\nabla \times \vec{a}),
$$
(known as ``kinetic'' helicity density when $\vec{a}=\vec{u}$, and
``magnetic'' helicity density when $\vec{a}=\vec{B}$, respectively \cite{Moffatt1978Magnetic}), and the
cross-helicity density
$$
\Hcross = \vec{u}\cdot\vec{B},
$$
all of which play important roles in the production of the magnetic
field by the chaotic convective flow.
\begin{figure}[ht]
\mypsfrag{Mdip}{\Mdip}
\mypsfrag{t}{$t$}
\mypsfrag{M}{$M$}
\mypsfrag{Ek}{$E$}
\mypsfrag{MD}{\MD}
\mypsfrag{FD}{\FD}
\mypsfrag{0}   {0}   
\mypsfrag{0.2} {0.2} 
\mypsfrag{0.4} {0.4} 
\mypsfrag{0.6} {0.6} 
\mypsfrag{0.8} {0.8} 
\mypsfrag{1.2} {1.2} 
\mypsfrag{1.6} {1.6} 
\mypsfrag{1000}{1000}
\mypsfrag{2000}{2000}
\mypsfrag{3000}{3000}
\mypsfrag{4000}{4000}
\mypsfrag{5000}{5000}
\mypsfrag{6000}{6000}
\mypsfrag{10000}{10000}
\mypsfrag{(a)}{(a)}
\mypsfrag{(b)}{(b)}
\mypsfrag{(c)}{(c)}
\mypsfrag{(d)}{(d)}
\mypsfrag{(e)}{(e)}
\mypsfrag{(f)}{(f)}
\vspace*{-3.5mm}
\epsfig{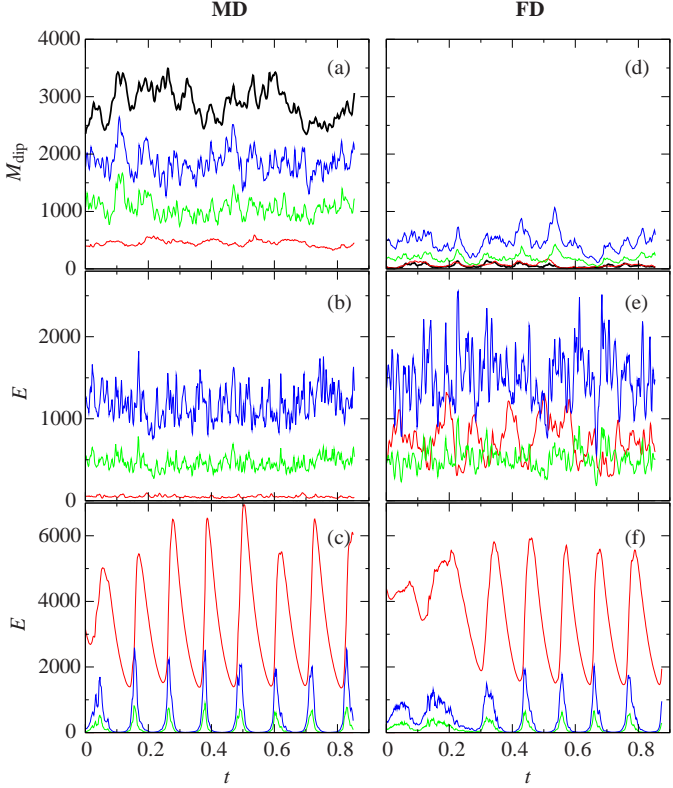}
\caption[]{(color online)
Coexisting distinct chaotic dynamo attractors at identical parameter
values -- a \MD (left column (a,b)) and a \FD dynamo (right column
(d,e)) both at $R=1.5\times10^6$, $\tau=2\times10^4$,
$P=0.75$ and  $P_m=1.5$.  Panels (a,d) show time series of
magnetic dipolar energy densities.
The rest of the panels show kinetic energy densities in the presence
of magnetic field (b,e) and after the magnetic field is removed
(c,f). The component $\overline{X}_p$ is shown by solid black
line, while $\overline{X}_t$, $\widetilde{X}_p$, and $\widetilde{X}_t$
are shown by red, green and blue lines, respectively. $X$ stands
for either $M$ or $E$.}  
\label{fig01}
\end{figure}
\begin{figure}[ht]
\mypsfrag{MfpMmp}{\MfpToMmp}
\mypsfrag{FDMD}{$X^{^\FD}/X^{^\MD}$}
\mypsfrag{ta}{\ttau}
\mypsfrag{10}{10}
\mypsfrag{100}{\hspace*{-8mm}$100 \times 10^3$}
\mypsfrag{1}{1}
\mypsfrag{(a)}{(a)}
\mypsfrag{(b)}{(b)}
\epsfig{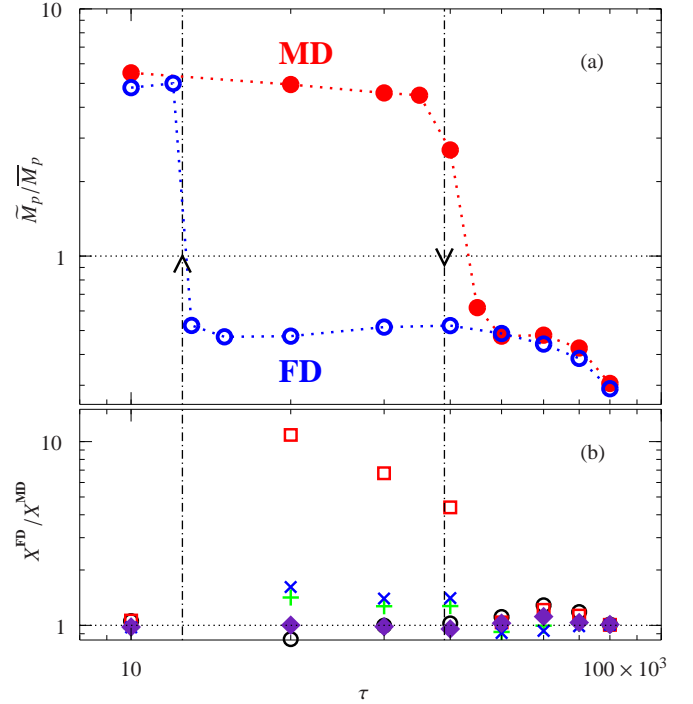} 
\caption[]{(color online)
 (a) Coexistence and hysteresis phenomena shown by the time-averaged ratio
 \MfpToMmp as a function of \ttau. \FD and \MD dynamos are
 indicated by full red and empty blue circles, respectively.
Expected locations of the transitions from \FD to \MD dynamos, and
vice versa are indicated by dash-dotted lines with arrows pointing
 down and up, respectively. 
(b) Comparison of convective properties of  \FD and \MD dynamos measured by
 the rations
$Nu_i^{_\FD}/Nu_i^{_\MD}$ (violet diamonds),
$\overline{E}_p^{_\FD}/\overline{E}_p^{_\MD}$ (black circles),
$\overline{E}_t^{_\FD}/\overline{E}_t^{_\MD}$ (red squares),
$\widetilde{E}_p^{_\FD}/\widetilde{E}_p^{_\MD}$ (green plus signs),
$\widetilde{E}_t^{_\FD}/\widetilde{E}_t^{_\MD}$ (blue crosses), all as
functions of \ttau.
Parameter values are $\P=0.75$, $\Pm=1.5$ and
 $\R=(5-3\cdot10^{-5}\ttau)\,R_c$, i.e.\ $\R\cdot
 10^{-5}=7.6,17,26,35,43,51,58,62$  at $\ttau=n \times 10^4,\; n=1,\dots8$.
}
\label{fig02}  
\end{figure}

\section{Bistability}

Typical examples of solutions to the dynamo problem outlined above
are shown in Figure~\ref{fig01}. The figure presents the main 
magnetic and kinetic energy density components of two distinct dynamo
cases as functions of time, and illustrates well the chaotic
nature of the solutions. Apart from the obvious quantitative
difference, an essential qualitative change in the balance of magnetic
energy components can be observed. The axisymmetric poloidal component
$\overline{M}_p$  is dominant in the case shown in Figure
\ref{fig01}(a,b) while it has a relatively small contribution in the
case of Figure \ref{fig01}(d,e).
This observation is in agreement with the claim made in
\cite{Kutzner2002From,Simitev2005Prandtlnumber,Christensen2006Scaling,Busse2006Parameter,Simitev2009,Olson2011} 
that, in general, two regimes of dipolar dynamos can be
distinguished, namely those with $\widetilde{M}_p < \overline{M}_p$
(denoted by \MD, "Mean Dipole" in \cite{Simitev2009} and below)
and those with $\widetilde{M}_p > \overline{M}_p$ (denoted by \FD,
"Fluctuating Dipole" in \cite{Simitev2009} and below). The dynamos
in Figure \ref{fig01}(a,b) and \ref{fig01}(d,e) are examples of 
these two types. A convenient measure of the type of dynamo is
therefore provided by the ratio $\widetilde{M}_p/\overline{M}_p$, which
we use extensively below.

Far more remarkable is the fact that the two distinct solutions shown
in Figure \ref{fig01} are obtained at identical parameter values and
coexist in this case. In fact, this is far from being an isolated
example. Indeed, varying the value of $\tau$ we find an extended
region of coexisting \MD{} and \FD{} dynamos as illustrated in
Figure \ref{fig02}(a) where the ratio
$\widetilde{M}_p/\overline{M}_p$ is plotted as a function of $\ttau$
in the case $P=0.75$, $\P_m=1.5$ and $R\approx 4\times R_c$, where $R_c$ is
the critical Rayleigh number for the onset of convection.
The transition between the \MD{} and \FD{} dynamos is discontinuous
and it is achieved via a hysteresis loop in the following sense. When
an \MD{} dynamo is used 
as initial data and the Coriolis number $\ttau$ is gradually
decreased, solutions remain in regime \MD{} until the critical value
\tB$\approx 12500$ is reached at which point an abrupt jump transition
to the \FD{} regime occurs. Similarly, when a \FD{} dynamo is used as
an initial condition and $\ttau$ is gradually increased the reverse
transition occurs at the critical value \tA$\approx 39000$ as seen in
Figure \ref{fig02}(a).  
The solutions plotted in Figure \ref{fig02} have been typically continued at
least up to 4 magnetic diffusion times. No evidence for a transient
nature in any case has been found. In fact, in cases outside of the
region of double attractors it takes typically less than 0.15 magnetic diffusion
times to switch from the initial conditions used to the appropriate
unique attractor. 

The results shown in Figure \ref{fig02}(a) are an important complement
to the findings of \cite{Simitev2009,Busse2010} where coexistence and
hysteresis was established as a function of the remaining
non-dimensional parameters $\P$, $\P_m$ and $R$ for fixed values of
$\ttau= 3\times 10^4,\; 4\times 10^4$ (see Figures 5 and 10 of
\cite{Simitev2009}, and \cite{Busse2010}, respectively). 
\begin{figure}[t]
\epsfig{file=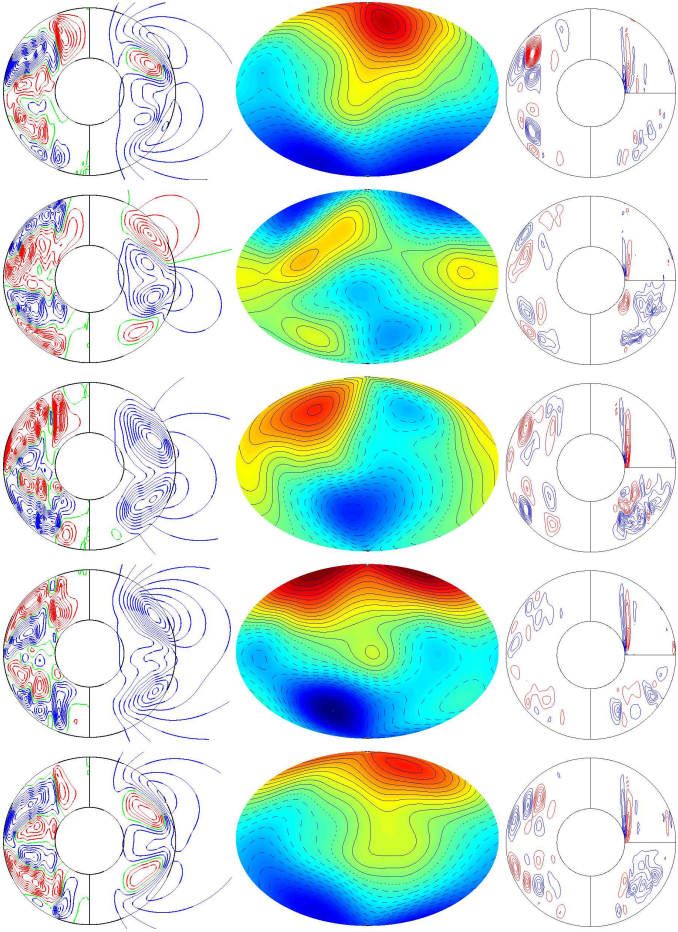,width=\columnwidth,clip=}
\caption[] {(color online)
A period of dipolar oscillations of a \FD{} dynamo. 
The plots in the leftmost column are meridional cuts showing lines of constant
  $\overline{B_{\varphi}}$ in the left half and of $r \sin \theta
  \dd_\theta \overline{h}$ in the right half. The plots in the middle
  column show  lines of constant $B_r$ at $r=r_o+1.3$. The plots in
  the rightmost column are 
  meridional cuts showing lines of constant densities of cross-helicity $\Hcross$,
  kinetic  helicity \Hu and magnetic helicity \HB
  at $\varphi=0$, in the left half, right upper and right lower
  quarters, respectively.
The rows correspond to equidistant moments separated by $\Delta
t=0.0252$. The parameter values are the same as in Figure \ref{fig01}.
}
\label{fig04}  
\end{figure}

\section{Comparison~of~bistable~attractors:~Magnetic~features}

All solutions included in Figure \ref{fig02} have a predominantly 
dipolar character with the ratio 
$\overline{M}_p^{\mathrm{dip}}/\overline{M}_p$, quantifying dipolarity
the field visible outside of the spherical shell, in the ranges
$[0.93, 1]$ for \MD{} dynamos and $[0.42, 0.70]$ for \FD{}  
dynamos, respectively. Although, the \FD{} dynamos feature an increased contribution
of higher multipoles they are of significant geophysical relevance
\cite{Kutzner2002From,Busse2006Parameter,Olson2011}.  
In this section we wish to discuss some of the magnetic properties of
\MD{} to \FD{} dynamos in more detail than it has been done previously
in \cite{Simitev2009,Busse2010}.

\subsection{Time dependence}  

Convection-driven dynamos exhibit chaotic time dependence (see
e.g.~Figure \ref{fig01}) except in
simple cases close to the critical value of $R$ and for rather large
values of $P_m$, such as the well-known dynamo benchmark case
\cite{benchmark}. There are, however, some coherent temporal features that 
can be distinguished. In particular, dynamos in the \FD{} regime are
typically oscillatory in that nearly periodic changes in
amplitude and field structures can be observed. As an illustration, Figure
\ref{fig04} shows a period of one such dipolar oscillation of
the \FD{} dynamo discussed in connection to Figure \ref{fig01}(d,e). 
At the beginning of the cycle magnetic flux with a ``new'' polarity is
generated near the equator of the inner boundary, and subsequently
propagates towards the poles replacing the flux of ``old'' 
polarity, as can be best seen in the plots of $\overline{B}_\varphi$
in this figure. The process repeats in a quasi-periodic
fashion. Kinetic helicity density remains largely unaffected 
while the magnetic and the cross-helicity densities participate in the
oscillation. 
On the other hand, dynamos in the \MD{} regime show 
significantly less variation in time and feature spacial structures  that
remain nearly static and fluctuate little with respect to their
time averages. This is due to the dominance of the mean components of
the poloidal magnetic field characteristic for the \MD{} regime.

\subsection{Spatial structures}  

\begin{figure}[t]
\epsfig{file=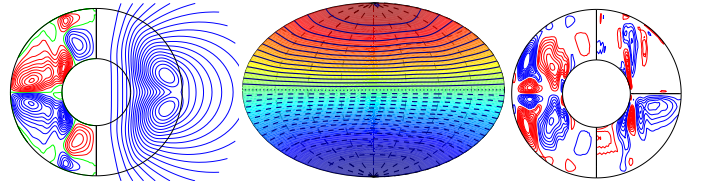,width=\columnwidth,clip=}
\caption[]{(color online)
  Time-averaged spatial structures of a \MD{} dynamo.  
  The same quantities are plotted as in Figure \ref{fig04}. The
  parameter values are the same as in Figure \ref{fig01}.
}
\label{fig03}
\end{figure}

Typical time-averaged spatial structures in the \MD{} regime are
illustrated in Figure \ref{fig03} by the example already discussed in
connection to Figure \ref{fig01}(a,b). The dynamo exhibits a nearly perfect and
relatively large-scale dipolar field best seen in the plots of the
radial magnetic field and the meridional field lines 
$r \sin \theta \dd_\theta \overline{h}=$ const. In particular, two
strong zonal magnetic flux tubes of  $\overline{B_{\varphi}}$ are 
formed inside the tangent cylinder, near the poles, while two tubes of
opposite polarity reside on both sides of the equator. The kinetic and
magnetic helicity densities are generated in narrow plumes primarily at the
boundary of the tangent cylinder, while the cross-helicity density forms
strong azimuthal tubes of alternating polarity filling the region
outside of the tangent cylinder.
In contrast, the spatial structures of \FD{} dynamos, which are
exhibited in Figure \ref{fig04}, have relatively
smaller scales and the evidence of higher multipole contributions are
clearly visible. At the minimum of the dipolar oscillation the zonal
magnetic flux tubes near the poles disappear, and the radial magnetic
field shows an excursion towards the opposite polarity  and an $m=1$
structure. The kinetic and magnetic helicity densities are again generated near
the tangent cylinder but appear smaller-scale and more fragmented. 


\section{Comparison~of~bistable~attractors:~Convective~features}

\subsection{Non-magnetic convection}

\begin{figure}[t]
\begin{center}
\epsfig{file=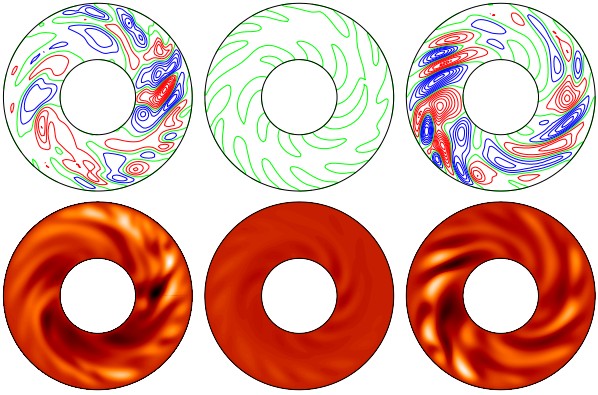,width=0.88\columnwidth,clip=}
\end{center}
\caption[]{(color online)
A period of relaxation oscillations seen in the equatorial plane of a
non-magnetic solution.  
The top row shows streamlines $r \partial_\varphi v=$ const and the
bottom row shows color maps of the temperature perturbation $\Theta$
both in the equatorial plane. The time step between plots is $\Delta
t=0.0448$. The same case is shown in Figure \ref{fig01}(f).  
}
\label{fig07}  
\end{figure}

The coexistence and hysteresis phenomena appear to be entirely
magnetic in nature. This is nicely illustrated in Figure 
\ref{fig01}(c,f) where the kinetic energy densities of two
non-magnetic solutions are shown, which are started from initial
conditions in the corresponding \MD{} and \FD{} dynamos in same
figure. Once the magnetic field is discarded, the flow quickly 
equilibrates to the same purely convective state. At these values of
the parameters, the convective state achieved is the well-known state
of relaxation oscillations \cite{GroBu,Simitev2003}. This state of
convection is remarkable in itself as it has a coherent nearly
periodic behaviour in an otherwise chaotic regime. Evidence of
its quasi-periodicity can be seen both in the time series of Figure
\ref{fig01}(c,f) as well as in the period of relaxation oscillations
shown in Figure \ref{fig07}.

Relaxation oscillations are one of a number of states achieved by
turbulent convection with increase of the Rayleigh number $R$. In this
state the differential rotation generated by the Reynolds stresses of
the convection columns becomes so large that it is able to destroy all
convective structures by shearing them off in the azimuthal direction.
In the absence of convection the differential rotation must decay since
there are no Reynolds stresses to sustain it. As the shearing action of the
differential rotation becomes sufficiently weak convection columns
grow in amplitude again. But as their Reynolds stresses regenerate the
differential rotation, their amplitude quickly peaks and then decays
as the shearing action interrupts the convection flows. It is
surprising how nearly periodically this process repeats itself even
though every convection episode differs from the next one in
detail. The period of these relaxation oscillations is primarily
governed by the viscous decay of the differential rotation. Since our study is based on the viscous decay time the value
of about 0.1 was found for the period over a wide range of the 
parameters $R$, $\tau$, and $P$. 

\subsection{Magnetic convection}

\begin{figure}[t]
\epsfig{file=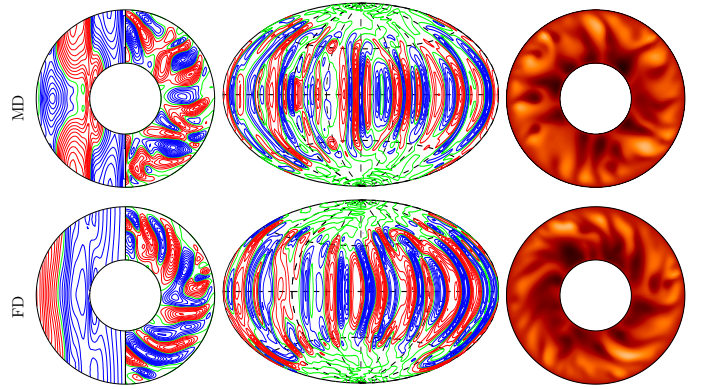,width=\columnwidth,clip=}
\caption[]{(color online)
Spatial structures of convection of a \MD  and a \FD dynamos.
The leftmost plots show lines of constant $\overline{u}_\varphi$ in the
meridional plane on the left, and streamlines $r \partial_\varphi
v=$ const.~in the equatorial plane on the right. The middle plots show
lines of constant $u_r$ on the spherical surface $r=r_i+0.5$. The
rightmost plots show color maps of the temperature perturbation
$\Theta$ in the equatorial plane. 
The parameter values are the same as in Figure \ref{fig01}.
}
\label{fig06}  
\end{figure}

The properties of convection in the presence of magnetic field differ
relatively little between dynamos in the \FD{} and \MD{} regime. The time
dependence of the respective convective flows can be compared in
panels (b) and (e) of Figure \ref{fig01} where kinetic energy density
components are shown. The time dependence is chaotic in both
cases, and it may be noted that convection is not in the state of
relaxation oscillations characteristic for the corresponding
non-magnetic cases. While the average values of the fluctuating
components \Efp and \Eft are nearly the same for the \FD{} and \MD{}
case, the main difference is the significant increase in differential rotation
measured by \Emt, which in the \FD{} dynamo is about 20 times larger
than in the \MD{} dynamo. Spatial structures of convection for the two
dynamos are shown in Figure \ref{fig06}. The length scales and the level
of irregularity of the structures is very similar for the two
regimes. The main difference appears in the profiles of the zonal
flow, which is nearly constant in $z$, the direction parallel to the axis
of rotation, in the \FD{} dynamo and strongly dependent on $z$ in the
\MD{} case. In addition, while the differential rotation near the
equator is in the prograde direction in the former case, it is in the
retrograde direction in the latter one. It is well-known that the main
effect of self-sustained magnetic field on convection is to strongly
inhibit differential rotation,
e.g.~\cite{Simitev2005Prandtlnumber}. This effect is stronger in the 
case of \MD{} dynamos which are characterized by stronger magnetic
fields and explains the observed differences.  
Evidence that these differences are typical throughout the \MD{} and
\FD{} regimes is presented in Figure \ref{fig02}(b) where ratios of
various kinetic energy components as well as the Nusselt numbers are
plotted for a number of coexisting attractors as a function of $\ttau$, and it
can be seen that their values do not differ significantly from unity
except in the case of $\Emt^{FD}/\Emt^{MD}$.
\begin{figure}[t]
\mypsfrag{M}{\Mdip}
\mypsfrag{t}{$t$}
\mypsfrag{0}   {0}   
\mypsfrag{0.5} {0.5} 
\mypsfrag{1}   {1} 
\mypsfrag{2} {2} 
\mypsfrag{1m4} {\hspace{1mm}$10^{-4}$} 
\mypsfrag{1m2} {\hspace{1mm}$10^{-2}$} 
\mypsfrag{1}   {\hspace{-1mm}1} 
\mypsfrag{1p2} {\hspace{1mm}$10^2$} 
\epsfig{file=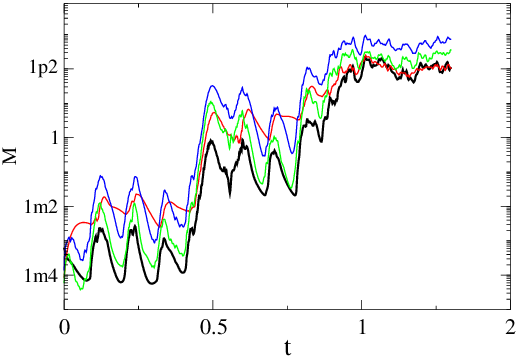,width=\columnwidth,clip=}
\caption[]{(color online)
Equilibration of a dynamo started from a small random magnetic field
seed. Time series of magnetic dipolar energy densities with the same
color scheme as in Figure \ref{fig01} are shown. The same parameter
values as in Figure \ref{fig01} are used.
}
\label{fig08}  
\vspace{2mm}
\mypsfrag{MfpMmp}{\MfpToMmp}
\mypsfrag{alp}{$\alpha$}
\mypsfrag{0}   {0}   
\mypsfrag{0.2} {0.2} 
\mypsfrag{0.4} {0.4} 
\mypsfrag{0.6} {0.6} 
\mypsfrag{0.8} {0.8} 
\mypsfrag{1.2} {1.2} 
\mypsfrag{1.6} {1.6} 
\mypsfrag{1}{1}
\begin{center}
\epsfig{file=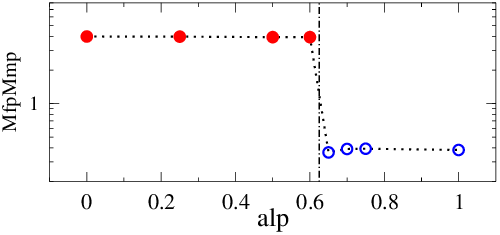,width=0.94\columnwidth,clip=}
\end{center}
\caption[]{(color online)
Width of the basins of attraction of \MD and \FD states shown by the
ratio \MfpToMmp as a function of the continuation parameter $\alpha$ 
of formula \eqref{cont}, in the case $R=1.5\times10^6$,
$\tau=2\times10^4$, $P=0.75$ and  $P_m=1.5$. The cases with $\alpha=0$
and $\alpha=1$ are the same \FD (red full circles) and the \MD dynamos
(blue empty circles) shown in Figure \ref{fig01}, respectively.
}
\label{fig09}  
\end{figure}

\subsection{Mechanism of coexistence}

The observations just made are useful in elaborating the mechanism of 
the coexistence phenomenon. 
Coexistence is the result of two different ways in which the magnetic
field damps the differential rotation to achieve the transport of the
same amount of heat.
In \MD{} dynamos the differential rotation generated by Reynolds
stresses of the convection columns is eliminated almost entirely by
the strong mean magnetic field.  Because of the strong magnetic field
the amplitude of convection is also reduced in comparison with the
maximum value that it reaches in the absence of a magnetic field. In
the case of \FD{} dynamos the differential rotation is still 
diminished, but its alignment with coaxial cylindrical surfaces is
preserved. The amplitude of convection is now more strongly
fluctuating, but is larger on average than in the case of the \MD{}
dynamos. In this way \FD{} and \MD{} dynamos manage to carry very nearly
the same heat transport as is evident from Figure \ref{fig02}(b). This heat
transport by far exceeds the time average of the heat transport found
in the absence of a magnetic field. 

\section{Basins of attraction}

Usually a dynamo equilibrates to a unique state. In contrast, for a
dynamo inside the region of coexistence, the initial conditions
determine whether a \FD{} or a \MD{} state is approached, as discussed
in connection with  Figure \ref{fig02}(a). Here we attempt to estimate the
set of initial conditions  that leads to equilibration to a \MD{} or a
\FD{} state. 

\subsection{Random initial conditions}

Figure \ref{fig08} shows a numerical experiment in which random
magnetic perturbation of small amplitude is applied to fully
developed convection in the regime of relaxation oscillations, and
integration in time is continued for the velocity, temperature and
magnetic fields. After a relatively long transient period of magnetic
field growth, the dynamo approaches the \FD{} regime. In addition,
this simulation suggests that the existence of a third chaotic
attractor along with the \FD{} and the \MD{} states is unlikely.

\subsection{Controlled initial conditions}

The vast majority of dynamo solutions published in the literature are
started from previously equilibrated runs with similar parameter
values in order to minimize transient times. In the region of
coexistence a dynamo started from another \FD{} (\MD{}) dynamo
approaches a \FD{} (\MD{}) state. In order to estimate the width
of the basins of attraction of the two coexisting states, we report in
Figure \ref{fig09} a number of simulations started from initial conditions
prepared in the form, 
\begin{equation}
\label{cont}
x(r,\theta,\varphi) = \alpha x^{_\MD}(r,\theta,\varphi) + (1-\alpha) x^{_\FD}(r,\theta,\varphi),
\end{equation}
where $\alpha\in[0,1]$ is a continuation parameter, $x$ represents
any of the dynamical variables $\vec{u}$, $\vec{B}$ and $\Theta$, and
the superscripts indicate equilibrated \FD{} and \MD{} dynamo
solutions. When $\alpha=0$ this is equivalent to initial conditions
chosen in the \FD{} regime, when $\alpha=1$ corresponds to initial conditions
chosen in the \MD{} regime, and variation of $\alpha\in(0,1)$ allows
us to follow a continuous path between the two attractors. Figure
\ref{fig09} shows that for the
test case $P=0.75$, $R=1.5\times10^6$, $\ttau=2\times10^4$ and
$\Pm=1.5$ the transition between \FD{} and \MD{} regimes occurs at
$\alpha=0.625\pm0.25$. 

We wish to comment that while both Figures \ref{fig08} and
\ref{fig09} show some bias towards the \FD{} regime, this may be  
due to our inability to select a test case situated exactly
in the middle of the coexistence region.

%
%
%
%

\section{Concluding remarks}

We have considered in this study a minimal self-consistent model of
dynamo action generated by convection in rotating spherical fluid
shells. While the relatively thick spherical shells, the relatively
large values of the Coriolis parameter, and the relatively low values
of the Rayleigh number employed here are more appropriate to
the problem of geo- and planetary magnetism, the model is generic and
may be used to understand solar and stellar magnetism. For instance,
the periodic reversals characteristic for \FD{} dynamos are
reminiscent of the 11-year Solar cycle. In particular,
we have been concerned here with the possibility of coexistence of
two nonlinear attractors in the fully-developed chaotic dynamo regime,
and the hysteretic transition between them. These phenomena have been
noted previously in \cite{Simitev2009,Busse2010} and discussed there
in some detail. The following is a summary of the important points
made here.

\begin{description}
\item[(a)]
We have established the coexistence of two nonlinear attractors,
denoted by \MD{} and \FD{} dynamos in the above, over a significantly
large interval of values of the Coriolis number $\ttau \in
(12500,39000)$. The transition between them takes the form of a
hysteresis loop. These results fill a gap left in \cite{Simitev2009,Busse2010}
and demonstrate that coexistence occurs as a function of all basic
parameters in the model.
\item[(b)]
We have discussed in detail the contrasting properties characterizing 
dynamos in the \MD{} and \FD{} regimes, including differences in
temporal behaviour and spatial structures of both magnetic field and
convection. We include new results on quantities important in
mean-field dynamo theories of magnetic field generation such as the
kinetic, magnetic and cross-helicity density profiles in time-averages as well
during oscillations and reversals.  
\item[(c)]
We have investigated the question of the width of the basins of attraction
of the coexisting chaotic states. 
\end{description}

The coexistence of two distinct turbulent attractors is also a
phenomenon of general interest as it is relatively rare in fluid
dynamics and magnetohydrodynamics. Finally, the range of values of the
Coriolis number $\ttau$ 
where we have found coexistence constitutes a rather large subinterval
of the range currently accessible by numerical simulations. This
requires extra care when numerical results are interpreted.

\section*{Acknowledgements}
The research reported in the paper was performed in parts during the
authors' participation in the 2011 program ``Dynamo, Dynamical Systems
and Topology'' at NORDITA. The research of R.~Simitev has also been 
supported by the UK Royal Society under Research Grant 2010 R2.


\end{document}